%% file: main.tex
\documentclass[12pt]{iopart}

\usepackage{iopams} 

\usepackage{graphicx}
\usepackage{booktabs}
\usepackage{multirow}
\usepackage{xcolor}
\usepackage[numbers,sort&compress]{natbib}
\usepackage{url}
\usepackage[colorlinks=true,linkcolor=black,citecolor=blue,urlcolor=blue]{hyperref}
\usepackage{orcidlink}
\usepackage{listings}

\begin{document}

\title[Radiation-Hard Low-Latency ML on FPGAs]{Enabling Low-Latency Machine learning on Radiation-Hard FPGAs with hls4ml}

\author{%
  Katya~Govorkova$^{\orcidlink{0000-0003-1920-6618}}$ \\ \address{Massachusetts Institute of Technology (MIT), 77 Massachusetts Ave, Cambridge, MA 02139, USA} \ead{ekaterina.govorkova@cern.ch}
}
\author{%
  Julián~García~Pardiñas$^{\orcidlink{0000-0003-2316-8829}}$ \\ \address{Massachusetts Institute of Technology (MIT), 77 Massachusetts Ave, Cambridge, MA 02139, USA} \ead{julian.garcia.pardinas@cern.ch}
}
\author{%
  Vladimir~Lon\v{c}ar$^{\orcidlink{0000-0003-3651-0232}}$ \\ \address{Department of Experimental Physics, European Organization for Nuclear Research (CERN), 1211 Geneva 23, Switzerland} \ead{vladimir.loncar@cern.ch}
}
\author{%
  Sebastian~Schmitt$^{\orcidlink{0000-0002-6394-1081}}$ \\ \address{Massachusetts Institute of Technology (MIT), 77 Massachusetts Ave, Cambridge, MA 02139, USA} \ead{s.schmitt@cern.ch}
}
\author{%
  Victoria~Nguyen$^{\orcidlink{0009-0001-8191-0908}}$ \\ \address{Massachusetts Institute of Technology (MIT), 77 Massachusetts Ave, Cambridge, MA 02139, USA} \ead{vkn@mit.edu}
}

\author{%
    Marco~Pizzichemi$^{\orcidlink{0000-0001-5189-230X}}$ \\ \address{University of Milano-Bicocca (Italy), and \\ Department of Experimental Physics, European Organization for Nuclear Research (CERN), 1211 Geneva 23, Switzerland}
}

\author{%
    Loris~Martinazzoli$^{\orcidlink{0000-0002-8996-795X}}$ \\ \address{Department of Experimental Physics, European Organization for Nuclear Research (CERN), 1211 Geneva 23, Switzerland} \ead{loris.martinazzoli@cern.ch} 
}
\author{%
  Eluned~Anne~Smith$^{\orcidlink{0000-0002-9740-0574}}$ \\ \address{Massachusetts Institute of Technology (MIT), 77 Massachusetts Ave, Cambridge, MA 02139, USA} \ead{eluned@mit.edu}
}

\begin{abstract}

This paper presents an end-to-end demonstration of a viable, ultra-fast, radiation-hard machine learning (ML) application on FPGAs, which could be used in future high-energy physics experiments.  We present a three-fold contribution, with the PicoCal calorimeter, planned for the LHCb Upgrade II experiment, used as a test case. First, we develop a lightweight autoencoder to compress a 32-sample timing readout, representative of that of the PicoCal, into a two-dimensional latent space. Second, we introduce a systematic, hardware-aware quantization strategy and show that the model can be reduced to 10-bit weights with minimal performance loss. Third, as a barrier to the adoption of on-detector ML is the lack of support for radiation-hard FPGAs in the High-Energy Physics community’s standard ML synthesis tool, \texttt{hls4ml}, we develop a new backend for this library.  This new back-end enables the automatic translation of ML models into High-Level Synthesis (HLS) projects for the Microchip PolarFire family of FPGAs,  one of the few commercially available and radiation hard FPGAs.  We present the synthesis of the autoencoder on a target PolarFire FPGA, which indicates that a latency of 25 ns can be achieved. We show that the resources utilized are low enough that the model can be placed within the inherently protected logic of the FPGA.   Our extension to \texttt{hls4ml} is a significant contribution, paving the way for broader adoption of ML on FPGAs in high-radiation environments. 

\end{abstract}




\input{sections/introduction.tex}
\input{sections/related_work}
\input{sections/model.tex}
\input{sections/compression.tex}
\input{sections/hardware_impl.tex}
\input{sections/discussion.tex}
\input{sections/conclusion.tex}

\section*{Acknowledgements}
\input{sections/acknowledgements}

\section*{Code Availability}
\label{sec:code_availability}

The \texttt{hls4ml} backend developed for this work is publicly available on GitHub at \url{https://github.com/fastmachinelearning/hls4ml}. The pipeline to reproduce the results in this paper is publicly available on GitLab at \url{https://gitlab.cern.ch/egovorko/ecal-data-compression/-/tree/paper}.

\pagebreak
\section*{Appendix}



\subsection{Generated code example}
\label{app:generated-code-example}

In this section, we show the main C++ file that was generated by \texttt{hls4ml} when synthesizing the autoencoder we describe in the paper. 
The rest of the code is attached as auxiliary material to this paper and includes the utility needed to define the different types of layers. 
\begin{lstlisting}
#include <iostream>

#include "myproject.h"
#include "parameters.h"

void myproject(
    hls::FIFO<input_array_t> &input_layer_fifo,
    hls::FIFO<result_array_t> &layer4_out_fifo
) {
    #pragma HLS function top
    #pragma HLS interface control type(simple)
    #pragma HLS interface argument(input_layer_fifo) type(simple)
    #pragma HLS interface argument(layer4_out_fifo) type(simple)
    #pragma HLS function pipeline II(4)

    // hls-fpga-machine-learning insert load weights
#ifndef __SYNTHESIS__
    static bool loaded_weights = false;
    if (!loaded_weights) {
        nnet::load_weights_from_txt<weight3_t, 64>(w3, "w3.txt");
        nnet::load_weights_from_txt<bias3_t, 2>(b3, "b3.txt");
        loaded_weights = true;
    }
#endif
    #pragma HLS memory partition variable(input_layer) type(complete) dim(0)
    input_t input_layer[32];
    input_array_t input_layer_struct = input_layer_fifo.read();
    for (unsigned i = 0; i < 32; i++) {
        input_layer[i] = input_layer_struct.data[i];
    }

    // hls-fpga-machine-learning insert layers

    #pragma HLS memory partition variable(layer3_out) type(complete) dim(0)
    layer3_t layer3_out[2];
    nnet::dense_latency<input_t, layer3_t, config3>(
        input_layer, layer3_out, w3, b3
    ); 
    // encoded

    #pragma HLS memory partition variable(layer4_out) type(complete) dim(0)
    result_t layer4_out[2];
    nnet::relu<layer3_t, result_t, relu_config4>(layer3_out, layer4_out); 
    // encoded_quantized_relu

    result_array_t layer4_out_struct;
    for (unsigned i = 0; i < 32; i++) {
        layer4_out_struct.data[i] = layer4_out[i];
    }
    layer4_out_fifo.write(layer4_out_struct);
}

\end{lstlisting}

\subsection{Floorplan example}
\label{app:floorplan-example}
\begin{figure}[tbh]
   \centering
   \includegraphics[width=0.8\linewidth]{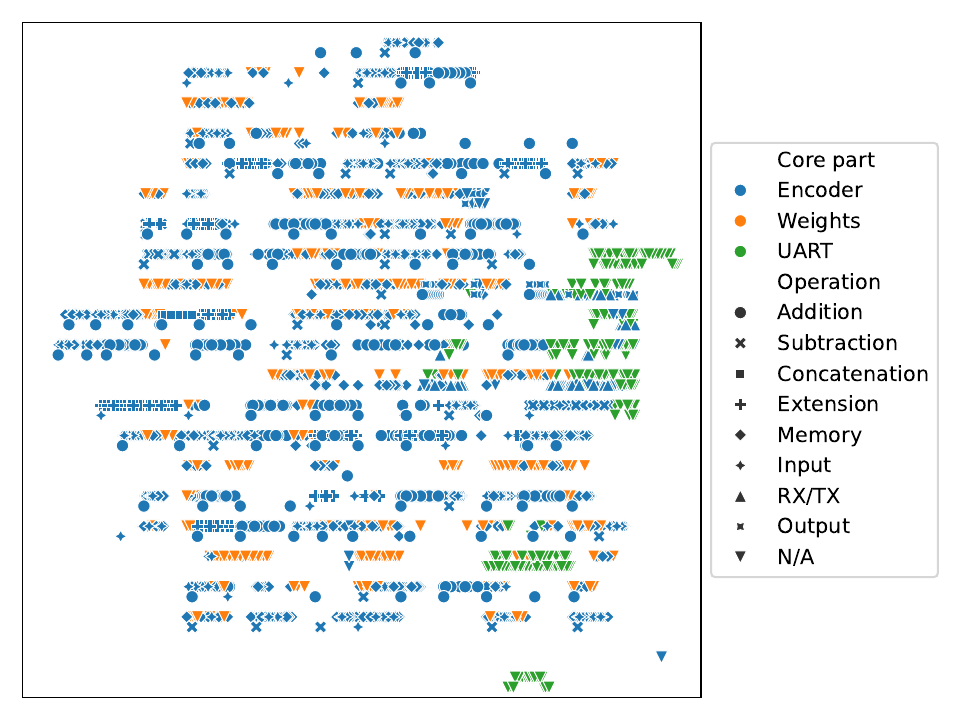}
   \caption{The floorplan of the nFPGA model. The core functionality includes the encoder, UART communication, and storage of weights and biases. For the encoder logic, separate markers are used to illustrate what operations are performed. N/A collects everything else from LED toggles to reset switches. The floorplan is zoomed in for visualization purposes.}
   \label{fig:floorplan}
\end{figure}

 In Fig.~\ref{fig:floorplan}, the floorplan of the nFPGA model is shown, where the x- and y-coordinates of the floorplan are used to illustrate the position of the wires and registers. 
The core functionality of the implementation, which the different parts are responsible for, is divided into UART communication, the core encoder, and the storage of weights and biases. For the encoder logic, separate markers are used to illustrate what operations are performed.  
For illustration purposes, this floorplan is heavily zoomed in and does not depict the full chip. 

\newpage

\bibliography{references}



\end{document}

%% file: sections/introduction.tex
\section{Introduction}
\label{sec:introduction}

The pursuit of new physics at the frontiers of energy and intensity requires constant innovation in detector technology and real-time data processing. The upcoming High-Luminosity Large Hadron Collider (HL-LHC) era is set to deliver an unprecedented volume of data~\cite{HL-LHC-TDR}. This increase in discovery potential is accompanied by significant challenges, most notably a massive increase in data rates and a higher number of simultaneous proton-proton collisions (pile-up). To handle this growing volume of data, experiments are increasingly forced to deploy powerful computational solutions as close to the detectors as possible. This ``edge computing" strategy minimizes data transmission latency and bandwidth but necessitates that compression be performed on electronics in harsh radiation environments \cite{lhc-rad-effects}. Consequently, there is a pressing need for ultra-fast, low-latency, and radiation-hard machine learning (ML) applications that can intelligently filter and compress data at the extreme edge of the readout chain.

The Large Hadron Collider beauty (LHCb) experiment's Upgrade II, planned to take place during the fourth Long Shutdown of the LHC, provides a compelling and concrete example of this challenge \cite{LHCb-Upgrade-II-Physics-Case}. This upgrade is designed to operate at a significantly higher instantaneous luminosity than before of up to $1.5\times10^{34}$~cm$^{-2}$~s$^{-1}$, which will increase the average pile-up to approximately 40. This higher collision rate, combined with the inherently large production cross-sections for the beauty and charm hadrons that LHCb targets, will generate an unprecedented data rate of 200 Tb/s, a volume that necessitates significant detector enhancements. An example of such a detector enhancement is the planned, high-granularity electromagnetic calorimeter with excellent timing capabilities, known as the PicoCal \cite{LHCb-Upgrade-II-Framework-TDR}. To achieve its target time resolution of $10-20$~ps, the SPIDER~\cite{alvado2025spiderwaveformdigitizerasic}, a waveform digitizer ASIC, will digitize the pulse shape from each calorimeter read-out channel. This pulse shape represents the time evolution of the voltage induced by the cascade of electrons produced within a photomultiplier tube (PMTs) \cite{LHCb-Upgrade-II-Framework-TDR} in response to scintillation photons generated by an electromagnetic shower in the calorimeter. This study assumes the pulse shape is represented by 32 16-bit numbers, although the final number of samples is yet to be determined, with current estimates ranging from 8 to 32.  The sampling interval is chosen such that the pulse rise time, defined as the time between the points where the pulse reaches 10\% and 90\% of its maximum amplitude, is captured by approximately four to five sampling points. This granularity is assumed to very roughly represent that expected of the LHCb PicoCal readout. However, it should be stressed that the performance of the machine-learning–based compression algorithms,  which is the focus of this paper,  should not depend critically on the precise granularity of the readout.
It is estimated that in order to have tenable data-rates between the on-detector front-end electronics and the backend electronics, the initial sample of 32 inputs must be compressed into a maximum of two numbers of a similar bit size on the detector. This work aims to achieve this compression ratio whilst still maintaining as much information about the pulse as possible. 
The hardware platform chosen for this design study is the Microchip PolarFire FPGA, a device selected for its inherent radiation tolerance. The technical motivations for this choice, which relate to its flash-based architecture as a superior alternative to traditional SRAM-based FPGAs requiring Triple Modular Redundancy (TMR), are detailed in Section~\ref{sec:radhard_paradigms}.

In high-energy physics, a common alternative to FPGAs is a custom Application-Specific Integrated Circuit (ASIC). However, the inflexible, multi-year development cycle and high non-recurring engineering costs mean ASICs are typically only used when extreme performance or power constraints cannot be met by programmable hardware. 
This work also serves to demonstrate that FPGAs can represent a viable alternative to ASICs for ML algorithms that must satisfy both demanding real-time requirements and run on-detector, in high-radiation environments. 

At the outset of this work, two key components for tackling the PicoCal data compression challenge were missing. First, no specific ML algorithm had been developed and validated for compression of the full pulse shape. Second, the high-level synthesis for machine learning (\texttt{hls4ml}) toolchain, a community standard for deploying ML models on FPGAs, lacked support for radiation-hard Microchip devices \cite{hls4ml-JINST}. This paper addresses these gaps with a three-fold contribution:
\begin{enumerate}
\item \textbf{An ML Algorithm for Pulse Compression:} We develop and validate a lightweight autoencoder model designed specifically for the LHCb PicoCal use case. We demonstrate through simulation that this model can effectively compress the 32-sample pulse shapes into a compact two-dimensional latent space, preserving the information required for downstream physics reconstruction.
\item \textbf{Hardware-Aware Model Compression:} We conduct a systematic study of model quantization, demonstrating a reduction in the model's computational complexity while preserving physics reconstruction performance.
\item \textbf{A New Backend for \texttt{hls4ml}:} We develop a new software backend for the \texttt{hls4ml} library~\cite{hls4ml-overview2} to support the Microchip SmartHLS compiler~\footnote{Microchip SmartHLS Tool Suite, Version 2021.2, Microchip Technology Inc., Chandler, AZ, USA.}. This critical infrastructure work enables, for the first time, the automated deployment of ML models onto radiation-hard Polar Fire FPGAs, opening the door for the wider high-energy physics community to leverage these devices for on-detector ML applications.
\end{enumerate}
The result is an end-to-end demonstration of a viable, ultra-fast ML application on a radiation-hard FPGA for a future high-energy physics experiment, a milestone facilitated by the new automated design flow also developed as part of this work.

For this design study, key performance and architectural parameters were established based on preliminary discussions with experts involved in the LHCb Upgrade II electronics design. A baseline scenario assumes that each front-end FPGA will need to process data from 8 independent calorimeter channels in parallel. Furthermore, while the LHC bunch crossing rate is 40~MHz, a target for the FPGA's internal processing clock of 160~MHz is considered a realistic benchmark. This choice allows for exactly four internal clock cycles to process each new input, setting a strict requirement for the initiation interval  of the compression algorithm. 

This paper is structured as follows. We first review related work in ML-based data compression and FPGA deployment toolchains in Section \ref{sec:related_work}. In Section~\ref{sec:model}, we describe the architecture of our autoencoder model and demonstrate its ability to compress and faithfully reconstruct pulse shapes while preserving key physics information. Section~\ref{sec:compression} details the hardware-aware quantization study performed to optimize the model for hardware implementation. In Section~\ref{sec:hardware_impl}, we present the development of the new \texttt{hls4ml} backend and the final synthesis results on the target PolarFire FPGA, showing that the design meets the application's strict performance requirements. Finally, we discuss the broader implications of these results in Section~\ref{sec:discussion} and present our conclusions in Section~\ref{sec:conclusion}.

%% file: sections/related_work.tex
\section{Related Work}
\label{sec:related_work}

This work builds upon established techniques in two distinct domains: the use of autoencoders for data processing in high-energy physics and the deployment of machine learning models on FPGAs using high-level synthesis tools. This section reviews the state-of-the-art in both areas to contextualize our contributions.

\subsection{Data Compression and Anomaly Detection with Autoencoders}
The immense data volume at the LHC has driven the widespread exploration of machine learning for real-time data reduction and analysis \cite{ml-for-hep}. Autoencoders, as a class of unsupervised neural networks, have proven to be particularly effective. They are trained to learn a compressed, low-dimensional representation (latent space) of high-dimensional data by reconstructing their own input, forcing the model to capture the most salient features of the data distribution.

In high-energy physics, this technique has been successfully applied to a range of problems. These include the compression of jet substructure information for trigger systems and offline storage, as well as model-agnostic anomaly detection, where events that are poorly reconstructed from the latent space are flagged as potential new physics \cite{Autoencoder-Anomaly-Detection-LHC}. Other studies have explored the physical meaning of the latent space itself, demonstrating that its learned variables can be correlated with underlying physical processes, forming a powerful feature space for analysis \cite{Autoencoder-Latent-Space-Physics}. Our work applies this paradigm to a novel and challenging use case: the on-detector, ultra-low-latency compression of full calorimeter pulse shapes, where preserving timing information is critical and preserving other pulse-shape information, such as the rise-time,  is desirable.

\subsection{ML-to-FPGA Toolchains and Radiation-Hard Hardware}
\label{sec:related_work_toolchains}
For ML models to be deployed at the detector front-end, they must meet stringent real-time processing constraints, with latencies typically ranging from nanoseconds to a few microseconds~\cite{hls4ml-JINST}. FPGAs are an ideal hardware platform for this task, enabling massively parallel and deeply pipelined implementations that eliminate external memory bottlenecks. The high-energy physics community has largely standardized on the \texttt{hls4ml} library for translating ML models from frameworks like TensorFlow/Keras into FPGA firmware. It has been successfully used to deploy a variety of models, from dense neural networks to recurrent networks, with latencies on the order of hundreds of nanoseconds. While commercial toolchains like Xilinx Vitis AI and Intel OpenVINO offer powerful solutions for deploying ML models on FPGAs, they are often tailored to specific vendor architectures and target a more general processing style with data and model residing in off-chip memory. In contrast, \texttt{hls4ml} real-time processing with models deployed fully on-chip is particularly well-suited to the needs of the scientific community, enabling fine-grained control over the hardware implementation and facilitating the integration of new backends, as demonstrated in this work. Furthermore, such automated workflows are crucial for enabling rapid design-space exploration, such as systematically varying model quantization, which would be significantly time-consuming with a purely manual HLS implementation.

However, \texttt{hls4ml} has historically supported HLS compilers exclusively for commercial FPGAs from Xilinx (now AMD) and Intel, which are typically SRAM-based. At the outset of this project, the lack of a streamlined, high-level toolchain for deploying ML models onto the radiation-hard, flash-based Microchip devices considered for this work was a  barrier to their adoption. This work closes that gap by developing and validating the first \texttt{hls4ml} backend for this class of FPGAs. 


\subsection{Radiation Hardness Paradigms in FPGAs}
\label{sec:radhard_paradigms}

Deploying electronics in the LHC's radiation environment necessitates robust mitigation against Single Event Effects (SEEs), which can disrupt circuit operation. For FPGAs, the mitigation strategy is intrinsically linked to the underlying technology of the configuration memory.

SRAM-based FPGAs, the most common commercial option, store their logic configuration in static RAM cells susceptible to SEUs, where a charged particle can corrupt the device's logic function. The standard mitigation technique is Triple Modular Redundancy (TMR), where logic is triplicated and a voter system corrects any single fault~\cite{tmr-fpga}. While effective, TMR imposes a significant overhead of at least 3x in logic resources, increased power consumption, and considerable design complexity.

In contrast, flash-based FPGAs, such as the Microchip PolarFire family used in this work, offer an alternative paradigm. Their configuration is stored in non-volatile flash memory cells, which are inherently immune to these configuration SEUs. This ``radiation-hardened-by-design'' approach avoids the need for configuration TMR, offering a path to reliable systems with significantly lower complexity and resource utilization. This fundamental difference was a key motivator for the hardware choice in this design study. The interplay between algorithm design, mitigation techniques, and the inherent robustness of the hardware is explored further in Section~\ref{sec:discussion}.

\subsection{Comparison with Related FPGA-Based ML Deployments}

Several recent works have demonstrated the use of \texttt{hls4ml} for instrumentation and data acquisition applications. For example, the EdgeML system developed for LCLS-II integrates machine learning inference within an FPGA-based data acquisition pipeline, achieving an inference latency of $\mathcal{O}(0.2\,\mu\mathrm{s})$ and a total system latency of $\sim 0.4\,\mu\mathrm{s}$~\cite{Mehdi_Rahimifar_2024}. Similarly, other studies have explored neural network deployment on MPSoC platforms using \texttt{hls4ml}-based workflows~\cite{auto_snl}, focusing on system-level integration and heterogeneous architectures.

In comparison, the present work targets a fundamentally different operating regime and hardware context. First, we focus on ultra-low-latency inference at the level of a single LHC bunch crossing, achieving a latency as low as $\sim 25\,\mathrm{ns}$, which is an order of magnitude lower than typical instrumentation-oriented implementations. Second, we target flash-based radiation-hard FPGAs (Microchip PolarFire), whereas most prior works rely on SRAM-based commercial devices without radiation constraints. Third, our approach enables fully on-detector deployment, where both model size and latency must satisfy strict front-end electronics requirements.

Finally, in contrast to existing \texttt{hls4ml} instrumentation efforts, this work introduces a new \texttt{hls4ml}  backend, faciltating automated deployment on radiation-hard devices.  

%% file: sections/model.tex
\section{Autoencoder Architecture and Co-Design Rationale}
\label{sec:model}

We aim to develop a lightweight autoencoder that achieves a high compression ratio while preserving the full pulse-shape information. Although extracting a single timestamp from the pulse shape is the most critical piece of information, retaining the full shape is highly desirable, as it provides additional insight into shower formation in the calorimeter and can help further mitigate pile-up effects.
The following subsections detail the autoencoder's specific architecture, its training, and its performance against key physics metrics. The algorithm performance presented in this section corresponds to results obtained after quantization-aware training, which motivated the choice to represent the latent space using two 10-bit values. The procedure used to implement and optimize this quantization-aware training is described in Section~\ref{sec:compression}.

\subsection{Model Architecture and Co-Design Rationale}
The autoencoder's architecture was co-designed with its hardware implementation in mind, prioritizing a minimal footprint and the lowest possible latency. The hardware-implemented portion is the encoder, which performs the on-detector data compression. It consists of two sequential layers:
\begin{itemize}
    \item A fully-connected dense layer that maps the 32 input data points directly to the two-dimensional latent space.
    \item A Rectified Linear Unit (ReLU) activation function is applied to the two latent space variables.
\end{itemize}
This involved directing the HLS compiler to create a parallel, pipelined architecture, relying on partially or fully unrolled loops, avoiding any resource sharing that would increase the processing time. 

For completeness, the decoder half of the autoencoder, used during training to reconstruct the pulse shape from the latent space, is constructed as a mirror of the encoder. It consists of a single fully-connected dense layer that maps the two latent space variables back to the 32-dimensional space of the output pulse shape. A linear activation function is used for this final layer to allow the output values to span the full, unconstrained range of the digitized pulse samples.

\subsection{Simulation Dataset}
The dataset used for training and evaluation consists of pulse shapes generated from a dedicated Monte Carlo simulation of the LHCb PicoCal electromagnetic calorimeter prototype~\cite{CERN-LHCC-2021-012}, implemented using the Geant4 toolkit~\cite{AGOSTINELLI2003250}. The simulation accurately models the detector geometry, material properties, and readout electronics to reproduce realistic signal responses under expected Run 5 conditions.

Two types of samples were prepared and then merged to mimic realistic detector conditions.
Signal pulses correspond to the responses of  SpaCal modules with a Pb absorber~\cite{CERN-LHCC-2021-012} to a single photon generated with energies uniformly distributed between $0.5 < E_T < 5~\mathrm{GeV}$, close to the proton-proton interaction point. Those are merged with background pulses, which were generated separately to represent pile-up and underlying event conditions. These were simulated assuming a fixed luminosity of $1.5 \times 10^{34}\ \mathrm{cm}^{-2}\,\mathrm{s}^{-1}$~\cite{CERN-LHCC-2021-012}. After generation, signal and background pulses were combined to emulate the complex conditions expected in LHCb Run 5 data-taking.

Each pulse was simulated with 1024 consecutive data points, reflecting the output of the DRS4-based V1742 CAEN digitizer (5 Gs/s, 500 MHz bandwidth)~\cite{BORDELIUS2025170608} used in the test beam setup. In the actual calorimeter, however, a different digitizer will be employed, providing only 32 samples per pulse. To match this configuration, we downsample the simulated waveforms from 1024 to 32 data points, with the rise-time of a typical pulse shape represented by around 1-3 sampling intervals. This is achieved by taking every 9 samples starting at the 380th to 668 out of 1024.

\subsection{Training}
We implemented fully connected autoencoders to reconstruct calorimeter pulse shapes represented as 32-sample vectors. The implementation was carried out using TensorFlow~\cite{tensorflow2015-whitepaper} with the Keras high-level API~\cite{chollet2015keras}. We train a full precision model as well as a quantized model for the FPGA implementation. The details about the quantization are covered in~Section~\ref{sec:compression}.

Training was performed using the Adam optimizer with a learning rate of $10^{-3}$ and the Mean Squared Error (MSE) loss function. Models were trained for up to 20 epochs with an early stopping criterion (patience of 5 epochs) based on validation loss to prevent overfitting. A batch size of 32 was applied during all experiments. The dataset of 353513 calorimeter pulses was split into 70\% training, 15\% validation, and 15\% test sets using a fixed random seed to ensure reproducibility. We used the "keras\_tuner"~\cite{chollet2015keras} random search to select the best learning rate and best model architecture by minimizing the MSE loss together with the number of FLOPS~(number of floating point operations per second). Since our input size was 32, we searched over the space of 15 combinations of keeping/dropping layers corresponding to sizes of 16, 8, 4, and 2 while maintaining that we always had at least one layer in the middle.  We converged on a 32-2 structure, as the MSE loss was approximately the same, and this was the smallest model with the smallest number of flops.  Before tuning the architecture, we tested using different activation functions (relu, elu, tanh), from which the relu activation function was found to be most performant. We additionally tested using different loss functions (Mean Squared Error, Mean Absolute Error, Huber, and Logcosh), but found no difference in performance.

\begin{figure}[tb]
  \centering
  \includegraphics[width=0.5\textwidth]{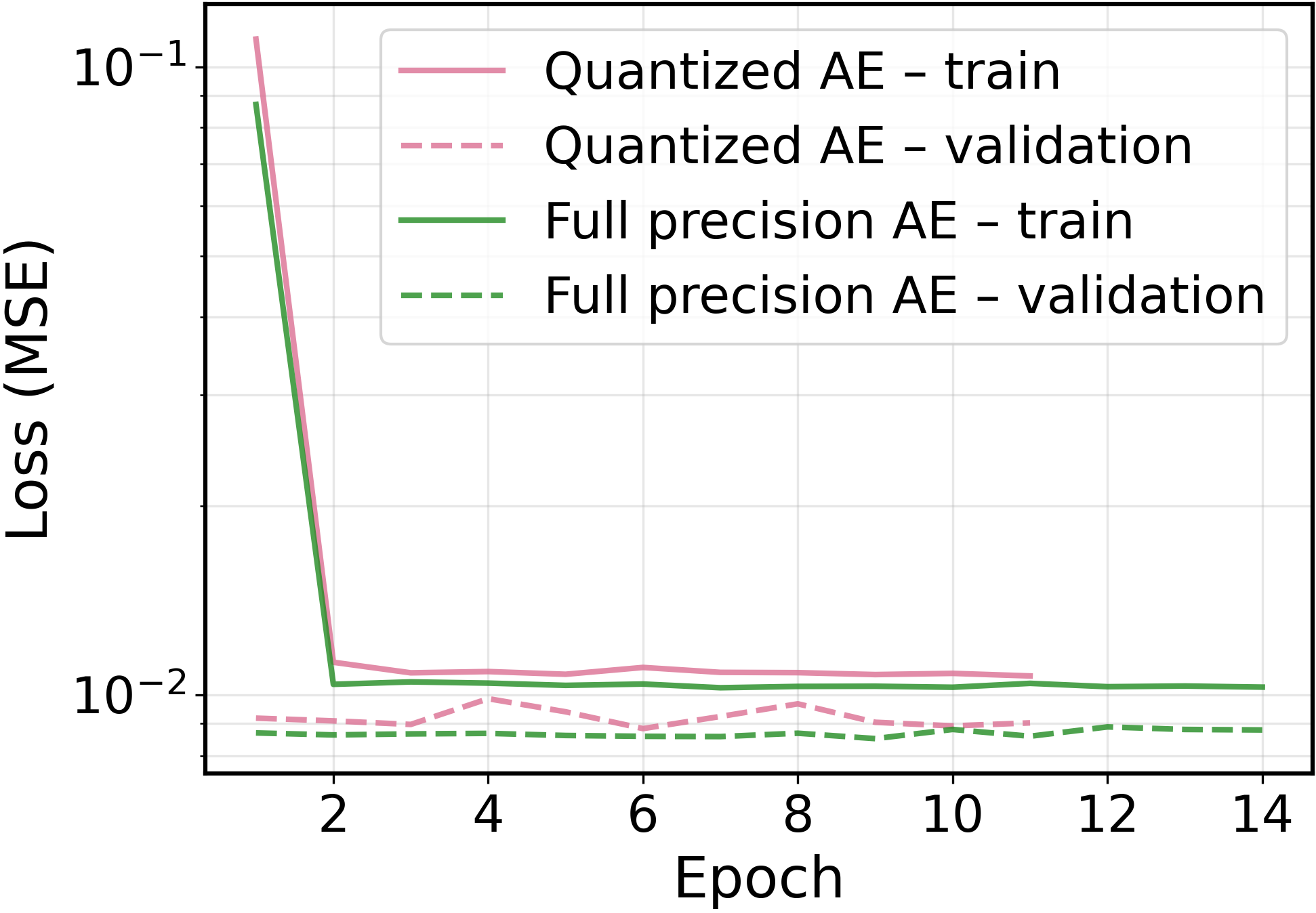}
  \caption{Training and validation loss curves for the full-precision autoencoder~(in green) and the quantized autoencoder~(in pink), discussed fully in Sec.~\ref{sec:compression}. The y-axis uses a logarithmic scale to emphasize convergence behavior.}
  \label{fig:training_loss}
\end{figure}

The training process was monitored using both training and validation loss, as shown in Figure~\ref{fig:training_loss}. The fact that the validation loss falls below the training loss is purely a statistical fluctuation, determined by the random seed used in partitioning the training and validation datasets. The best-performing weights, determined by the minimum validation loss, were restored at the end of training. After training, models were evaluated on the held-out test set using the MSE metric. 

\subsection{Pulse Shape Reconstruction Performance}
\begin{figure}[tb]
  \centering
  \includegraphics[width=\textwidth]{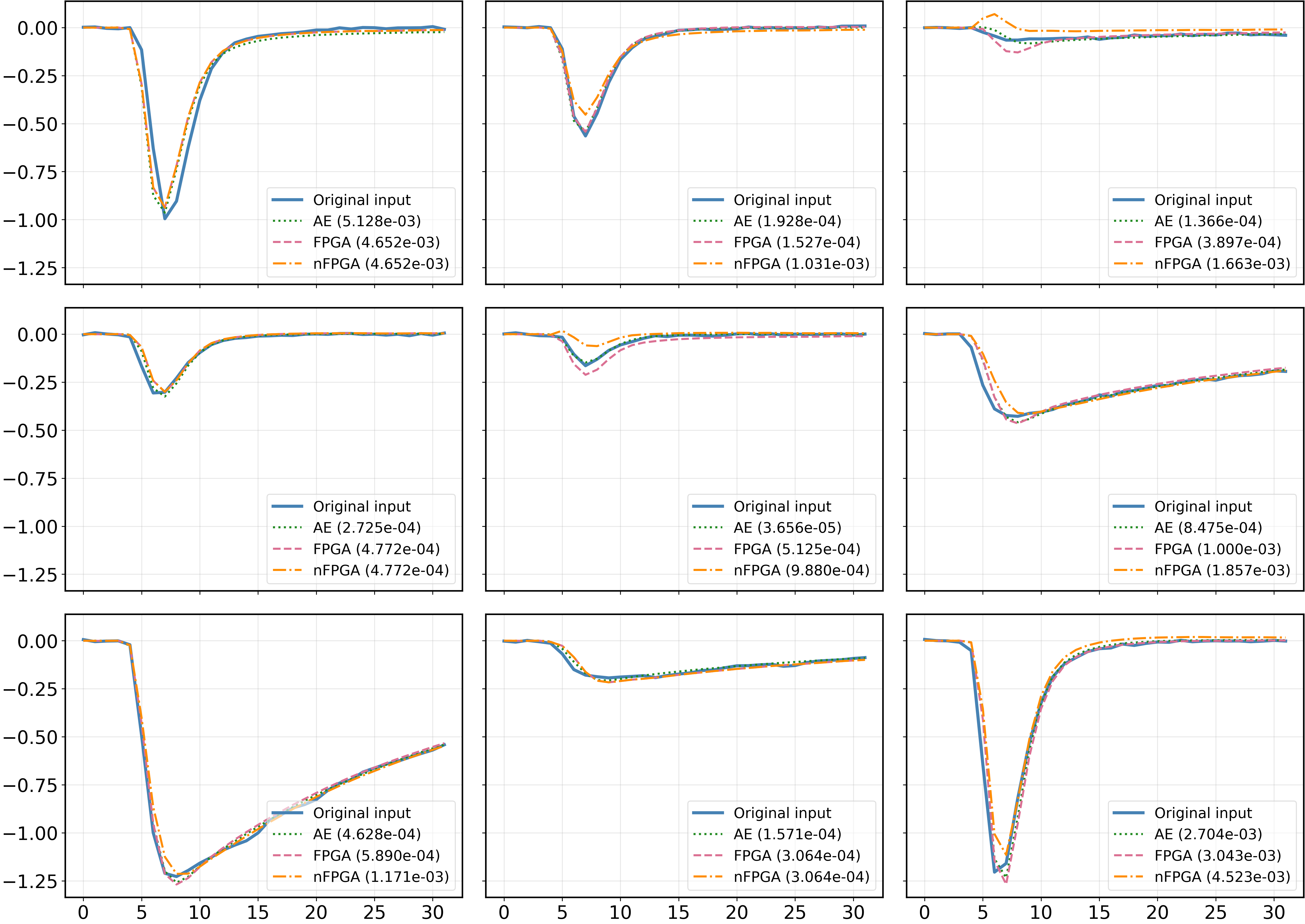}
  \caption{Examples of autoencoder reconstruction performance on calorimeter pulse shapes from the test set. Original waveforms (solid blue lines) are compared with their corresponding reconstructions (dashed/dotted lines). In green is the full precision model~(AE), in pink is the quantized model that produced the latent space running on the actual hardware~(FPGA). The orange line is also produced from a the latent space running on the actual hardware, but uses a more aggressively quantized model~(nano-FPGA), discussed in Sec.~\ref {sec:compression}. Visual agreement is supported by low MSE values across diverse pulse amplitudes, highlighting the robustness of the reconstruction.}
  \label{fig:reco}
\end{figure}

The performance of the trained autoencoder was initially assessed by visually comparing original input pulse shapes with their reconstructed counterparts and by evaluating the MSE on the test set. Figure \ref{fig:reco} presents several representative examples where the original waveforms (solid blue lines) are overlaid with reconstructions (dotted green lines) obtained with Python after compression into the 2-dimensional latent space. The pink dashed line shows the reconstruction using the latent space of a 10-bit precision encoder run on a PolarFire FPGA (full hardware details are given in Sec.~\ref{sec:hardwaredets}). This is referred to the `full' FPGA model.  The orange dashed-dotted line shows the reconstruction using a 10-bit precision encoder also run on an FPGA, but with  a lower precision used for the internal math. This referred to the nano-FPGA model, and is discussed further in Sec. 4. 

The close visual alignment between the original and reconstructed signals across a wide range of amplitudes confirms that the autoencoder effectively preserves both global and fine-grained features of the calorimeter response. This observation is consistent with the low average MSE values measured on the validation and test sets, demonstrating that the model achieves high-fidelity reconstruction even for challenging cases such as pulses with significant noise or extended tails.  The only exception is for very  low-energy pulses replicated by the quantized FPGA models, which occasionally under or over shoot the original lineshape. However these mis-alignments are for the most part small, and these lower-energy pulse will contribute relatively little to the total energy and timing estimate of the final cluster.  

\subsection{Latent Space Analysis}
\label{sec:latent_space}
To assess the interpretability of the learned latent space, we analyzed its correlation with key physical properties of the pulse shapes: \textit{true timestamp} (time of arrival of the simulated particle to the PicoCal), \textit{peak amplitude} and \textit{rising time}, 
where the latter provides a shape-sensitive indicator of the signal development.

\begin{figure}[t]
  \centering
  \includegraphics[width=\textwidth]{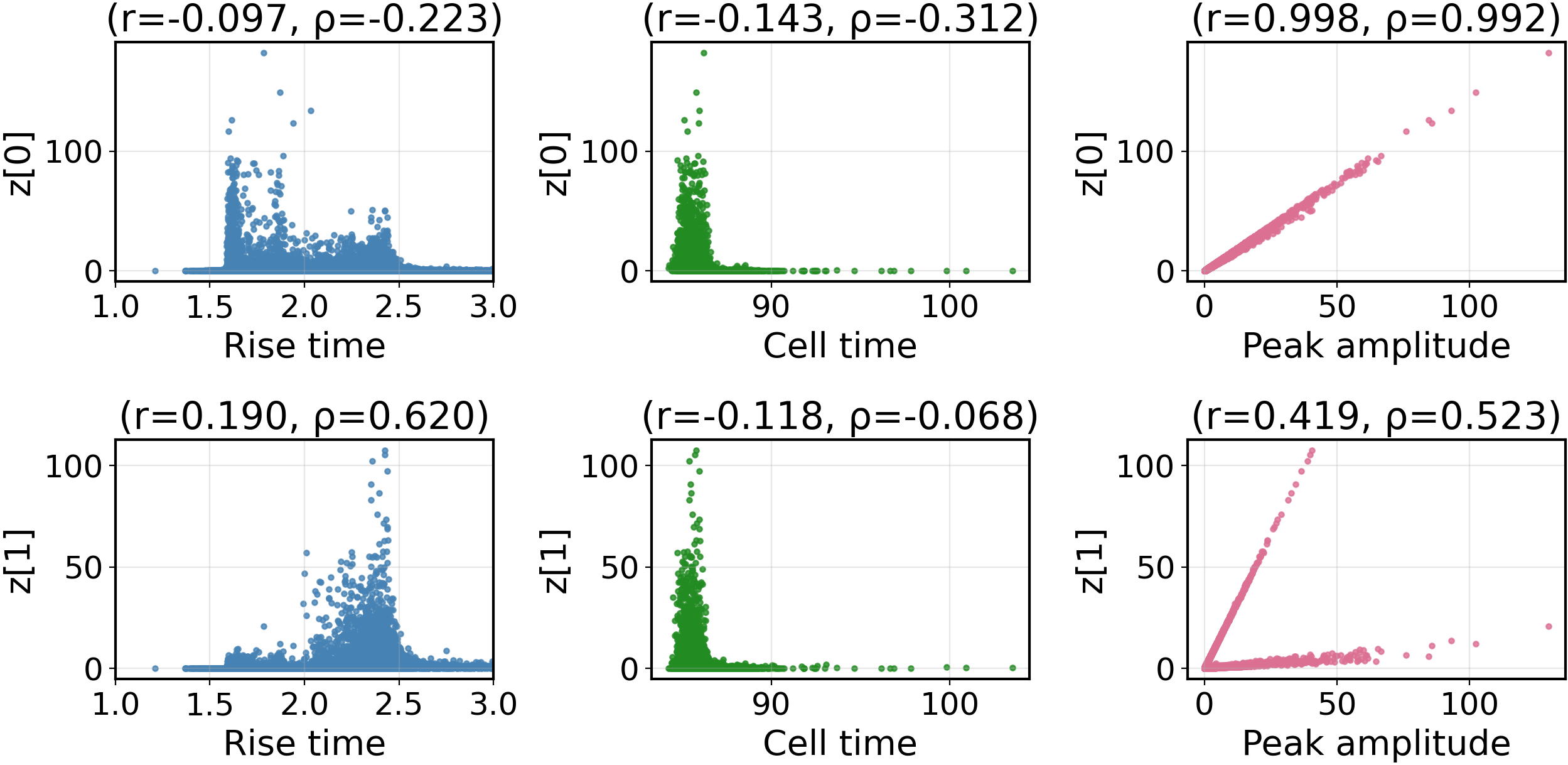}
  \caption{
  Correlation between latent space variables of the full-precision autoencoder and three pulse-level features on the test set: rise time (10\%--90\% interval), pulse true timestamp, and peak amplitude. Each row corresponds to one latent dimension ($z[0], z[1]$), and each column to one feature. Scatter plots include Pearson's $r$ and Spearman's $\rho$ coefficients in the titles. The latent representation is strongly correlated with peak amplitude, while correlations with timing features (rise time and cell time) are weaker.
  }
  \label{fig:latent-correlations}
\end{figure}

Figure~\ref{fig:latent-correlations} summarizes the Pearson and Spearman correlation coefficients between the two latent variables ($z_0$, $z_1$) and these physical quantities.


The results reveal a clear structure in the latent space. The first dimension ($z_0$) shows an almost perfect positive correlation with the \textit{peak amplitude} (Pearson $=0.998$, Spearman $=0.992$), indicating that this axis primarily encodes amplitude-related information. The second latent variable ($z_1$) also correlates with amplitude, suggesting it captures secondary shape-dependent features.

Both latent variables exhibit modest negative correlations with the \textit{true timestamp} (Pearson $\approx -0.12$ to $-0.14$) and the \textit{rising time} (Pearson up to $-0.1$). This suggests that timing and shape information are partially disentangled in the latent representation, although less strongly than amplitude.

Overall, these findings indicate that the autoencoder organizes the two-dimensional latent space with one dimension dominated by amplitude information, and another influenced by shape and timing variations. This interpretable structure increases confidence that the compression preserves the essential physics characteristics necessary for downstream tasks such as time reconstruction.

\subsection{Validation of Timestamp and Rise-time Reconstruction}
Time reconstruction, expressed as a timestamp, refers to determining the precise arrival time of a signal from the pulse shape it generates in the PMT. This is typically done using algorithms such as the Constant Fraction Discrimination (CFD)~\cite{Gedcke1968}. Preserving this information after compression is crucial, as it underpins the ability to correctly group cells originating from the same incoming particle in the high pile-up environment of LHCb Upgrade II, and to associate the particle with the correct proton–proton interaction~\cite{CERN-LHCC-2021-012}. 

As further validation of our autoencoder, we therefore wish to apply the standard CFD timing algorithm to both the original and the autoencoder-reconstructed pulses from the held-out test set, and compare both with the true timestamp. 

While injecting the underlying event into the baseline simulation produces a more realistic readout scenario, it obscures the true arrival time of a pulse, because it is no longer possible to determine whether a given pulse shape originates from an injected signal photon (for which the true time is known) or from other photons in the event (for which it is not). As a result, the timestamp reconstruction is tested on pulses from so-called seed cells only. To identify these, we perform a simple clustering of cells, selecting the one with the highest local energy (the seed cell). We then match clusters to the true simulated photons. Taking the seed cell from the matched cluster ensures that the arrival time corresponds to the simulated photon and not to the underlying event; therefore, we can use it to perform the time-extraction study detailed below. 
Additionally, two distinct types of pulses are observed due to the geometry of the spaghetti calorimeter (SpaCal), which is the detector type in the central region of the PicoCal~\cite{CERN-LHCC-2021-012}, composed of longitudinal scintillators intersected by a mirror along the longitudinal plane. Photons can either be reflected by this mirror and read out at the front of the SpaCal (front pulses) or they can originate beyond the mirror and be read out at the back of the SpaCal (back pulses). These two pulse-types have different offsets from the true arrival time, due to differing distances between the active material and the read-out, and so for the purpose of validation, we restrict the data-set to just one pulse-type (front pulses). It should be stressed, however, that it is still the baseline quantized autoencoder used in the validation, namely the FPGA model shown in pink in Fig.~\ref{fig:reco}, which is trained on both pulse-types and quantized to 10 bits.

The CFD algorithm estimates the signal timestamp by finding the point at which the pulse crosses a fixed fraction (20\%) of its maximum amplitude after baseline subtraction. The same CFD configuration was adopted for both original and autoencoder-reconstructed pulses to ensure a fair comparison. The reconstructed timestamps were then compared against the ground-truth timing, corresponding to the simulated time of arrival of the particle in the calorimeter. 

To account for a constant bias in the CFD estimator arising from the difference between the front pulse time of arrival and the true time, the mean residual obtained on the training sample was subtracted from all test results. This correction serves purely to remove a fixed offset, which would also be calibrated for in a real detector implementation. The constant offset was calculated on the training data for both the original and reconstructed pulses. 

\begin{figure}[tb]
    \centering
    \includegraphics[width=0.48\textwidth]{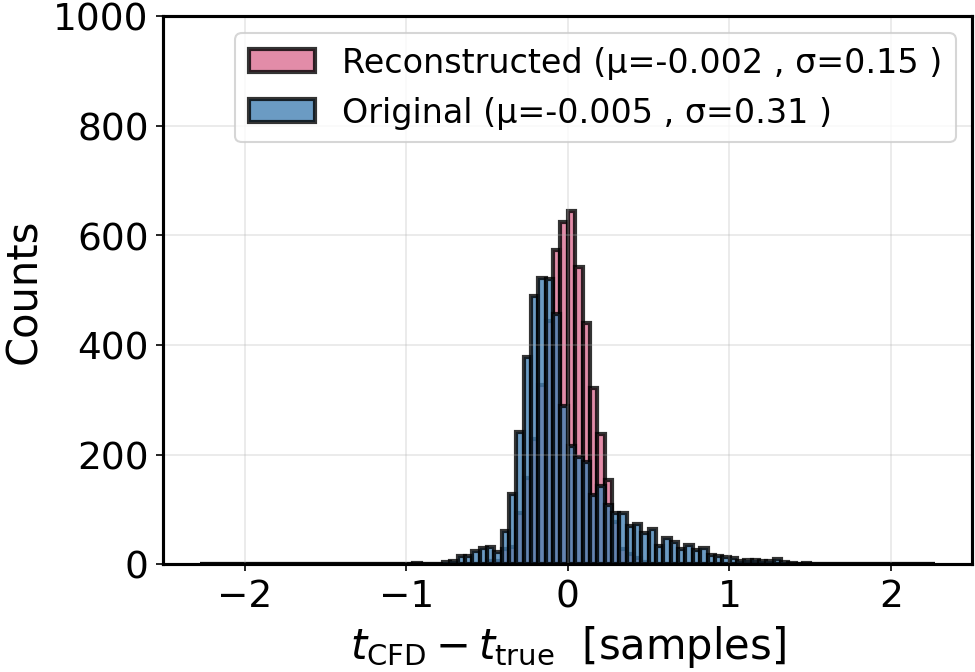}
    \includegraphics[width=0.48\textwidth]{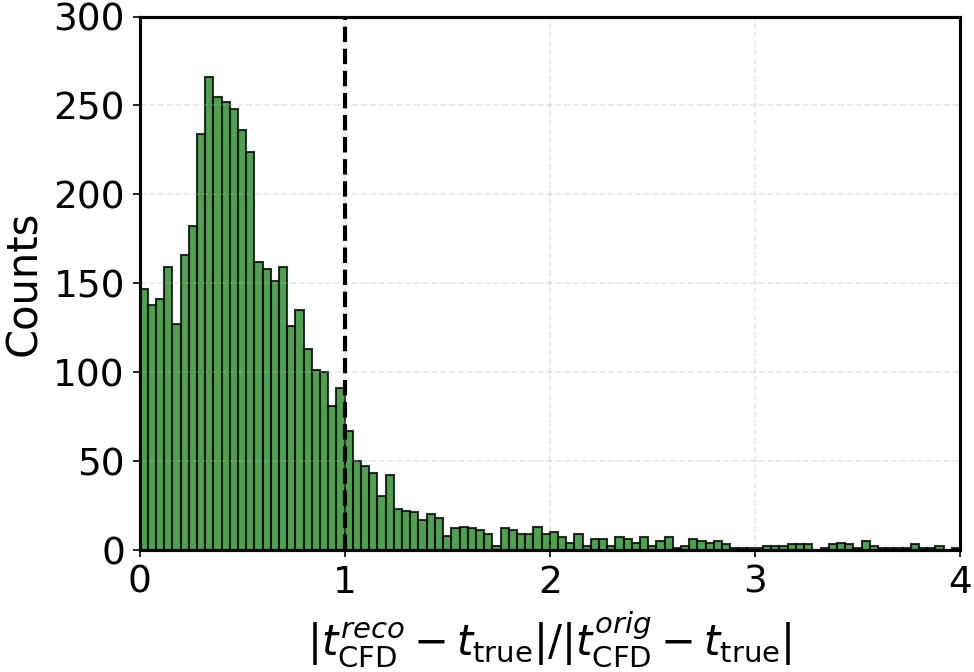}
    \caption{Residual distributions between CFD-reconstructed timestamps and true simulation times for original and autoencoder-reconstructed pulses~(left), and the ratio of their absolute values~(right). Reconstruction systematically reduces the timing residual for approximately half of the events.}
    \label{fig:cfd_residuals}
\end{figure}

It should also be noted that the comparison is performed between the ML-model run on an FPGA with quantized input and quantized encoder, while the CFD algorithm runs in full precision on full-precision input pulses. 
Figure~\ref{fig:cfd_residuals} shows the distributions of CFD timing residuals 
between reconstructed and simulated arrival times. The left panel compares the 
residuals obtained from the original 32-sample pulses and from the 
autoencoder-reconstructed pulses using identical CFD parameters. After applying the constant-bias calibration (mean subtraction estimated on the training set), both distributions are centred roughly around zero, as expected. The key comparison of interest is the width of the distributions. The reconstructed pulses exhibit a substantially narrower residual distribution than the original 32-sample pulses, indicating an improvement in timing precision of about a factor of two. Additionally, the right panel presents the ratio of absolute residuals for the reconstructed pulses over the absolute residuals of the original pulse. This quantifies the event-by-event performance gain, which is better than two (i.e. a ratio less than 0.5) for approximately half of the events.  
The units for the standard deviation shown in Figure \ref{fig:cfd_residuals} (left) are expressed in sampling intervals. Assuming the upper end of the expected sampling period for the readout—approximately 200~ps~\footnote{\url{https://indico.cern.ch/event/1502285/contributions/6554431/attachments/3081316/5592641/SPIDER_oral_TWEPP2025.pdf}} —this corresponds to a standard deviation of about 30 ps for the reconstructed pulses and 62 ps for the original pulses. However, it should be emphasized that the simulation used in this study does not incorporate the final version of the digitizer, and therefore, the approximate “downsampling” performed manually may not fully represent the characteristics of the final digitized output and, therefore, the final expected timing-resolution.   Similarly, the exact waveform characteristics can change with the final version of digitizer, which in turn could alter these results. 

Overall the results confirm that the autoencoder effectively denoises and restores the temporal structure of the pulses, and that is smoothing effect actually leads to a far more accurate and stable CFD-based timestamp reconstruction.


\begin{figure}[tb]
  \centering
  \includegraphics[width=0.5\textwidth]{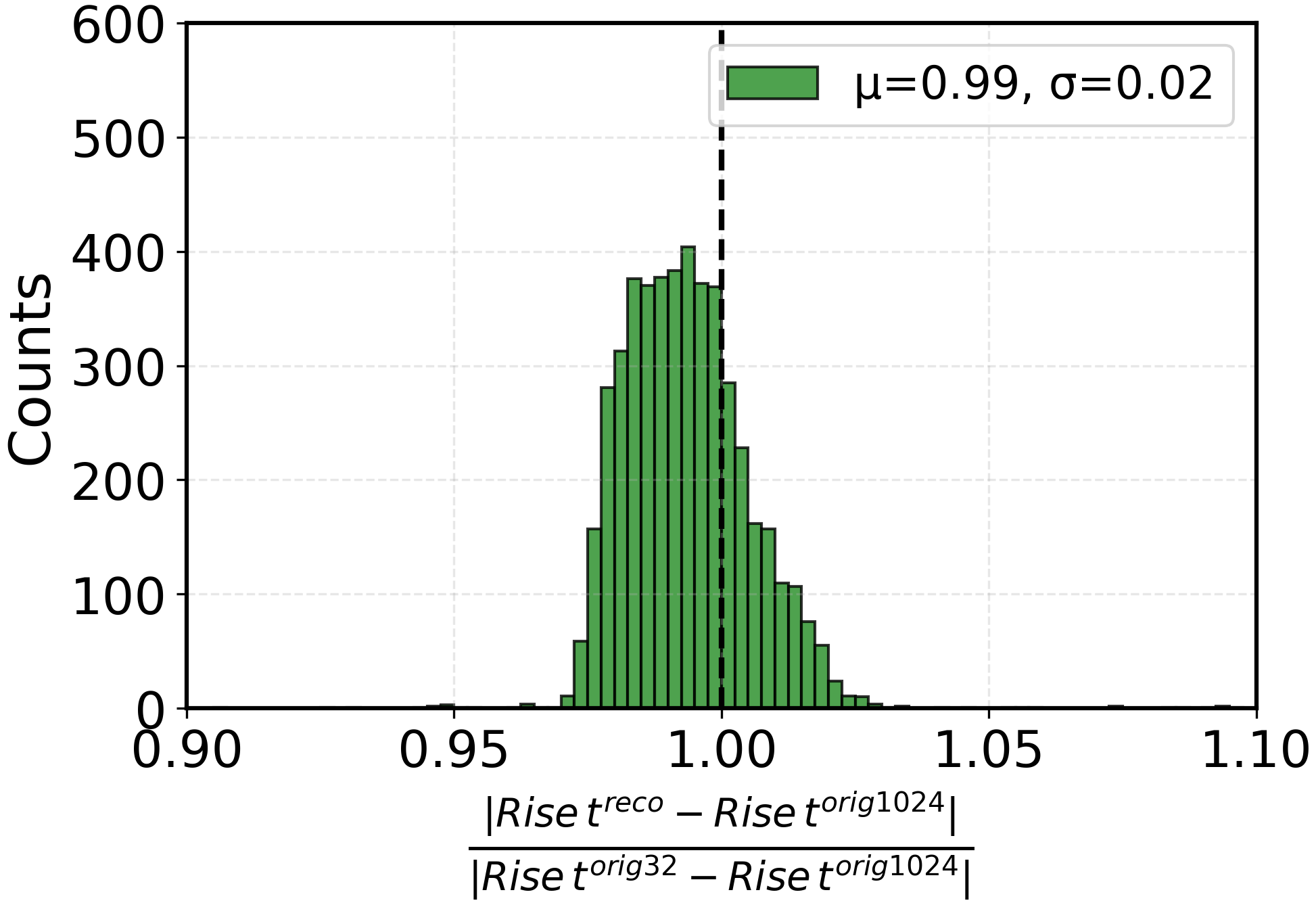}
  \caption{Histogram of the ratio of absolute differences in 10--90\,\% rise time: (reconstructed vs.\ 32-sample original) relative to (32-sample original vs.\ 1024-sample original).}
  \label{fig:rt-correlations}
\end{figure}

The histogram in Fig.~\ref{fig:rt-correlations} quantifies how accurately the reconstructed pulses reproduce the temporal characteristics of the original waveforms by comparing the deviation of their 10--90\,\% rise times from the "true" rise-time, which is taken as that found on the original 1024-sample pulses. Similarly to the CFD residuals, the mean bias in the rise-time difference due to downsampling from 1024 to 32 samples was subtracted separately for the original and reconstructed pulses using the training dataset before ratio evaluation. This ensures that the comparison focuses on the relative reconstruction precision rather than on the systematic offset introduced by coarser sampling. The autoencoder successfully restores fine temporal features lost during the 32-sample discretisation, achieving rise-time consistency with the full-resolution (1024-sample) reference within a few percent and again showing a modest increase in performance compared to the original 32-sample pulses.


\subsection{Timestamp Regression}
\begin{figure}[tb]
  \centering
  \includegraphics[width=0.49\textwidth]{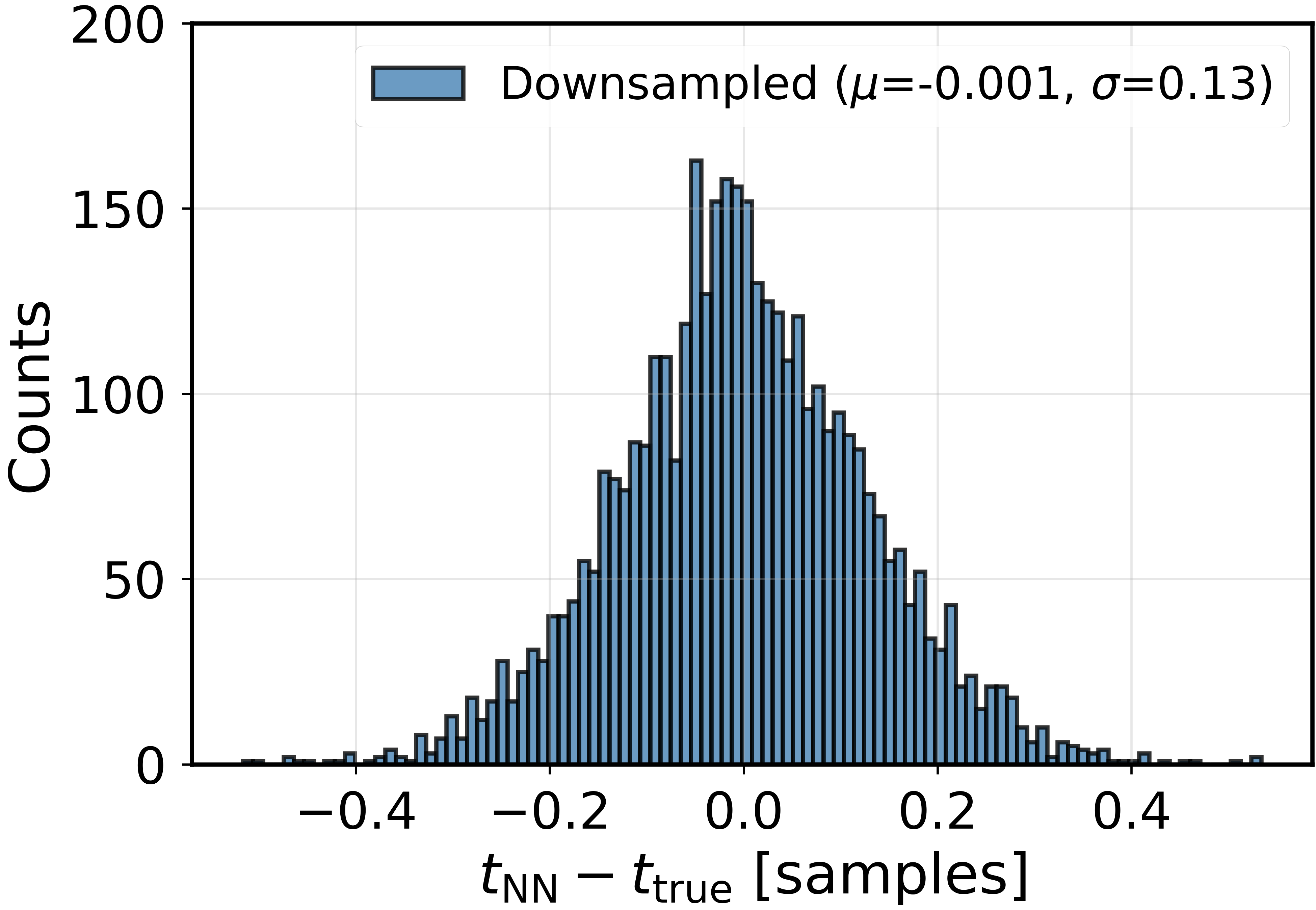}
\caption{Timestamp–residual distributions for the timestamps predicted using linear regression. The panel shows the histogram of $\Delta t = \hat{t} - t_{\mathrm{true}}$; the legend reports the mean $\mu$ and standard deviation $\sigma$. All values are in units of samples.}
  \label{fig:regression_residuals}
\end{figure}

To further validate our approach, we investigated whether a model with the same number of layers but trained to directly regress the timestamp from the 32-sample pulse (i.e.\ $32 \rightarrow 1$) could outperform the autoencoder-based reconstruction followed by the CFD evaluation. The network architecture was reinvestiagted for this set-up. It was found that a network containing no activation function in the output, i.e. a simple a linear regressor, was very performant. The model was trained jointly on both front and back pulses, in the same way as the baseline autoencoder in this paper, to ensure a fair comparison. However, marginally better performance is achievable if the model is trained separately. The residuals $(t_{\mathrm{pred}} - t_{\mathrm{true}})$, shown in Fig.~\ref{fig:regression_residuals}, were then analyzed.

These results indicate that the direct timestamp regression achieves a modest improvement over the default model ($\sigma = 0.13$ vs $\sigma = 0.15$).  However the 32–2 autoencoder also provides a compact, denoised pulse representation that can support other reconstruction tasks and aid in identifying anomalous signals such as pile-up–induced double peaks.
In either case, since this study demonstrates that a 32–2 model can be implemented within the available processing logic, reducing it to a 32–1 model would be straightforward. Future work should determine the level of compression that ultimately provides the optimal balance between performance and efficiency.



%% file: sections/compression.tex
\section{Hardware-Aware Quantization for Efficient FPGA Inference}
\label{sec:compression}

\subsection{Quantization and Impact Analysis}

To prepare the autoencoder for an efficient hardware implementation, the model’s inputs, activations, and weights were converted to fixed-point data types. A mixed-precision scheme was adopted to optimize hardware resource usage while preserving physics performance. The model inputs are represented using a 16-bit fixed-point format with 9 integer bits (including the sign bit), denoted as $<16,9>$. For the computational core, the dense layer’s weights and biases are quantized more aggressively to a 10-bit fixed-point representation with 4 integer bits (including the sign bit), $<10,4>$. The ReLU activation is quantized to $<10,7>$, which effectively means that the latent space is represented by two 10-bit numbers. The decoder is left to be trained and evaluated at full precision, as it is never going to be deployed on the FPGA itself.

The choice of weight precision is supported by the quantization scan shown in Figure~7.
During the quantization procedure, the internal weights, biases, and activations were quantized using the same total bit-width, while allowing for different allocations of integer bits. This reflects the differing dynamic ranges of these quantities: weights and biases are typically confined to a narrow range close to zero, whereas activations span a broader range and therefore require more integer bits to avoid saturation. In the final configuration, weights and biases are represented using $<10,4>$, while the latent space activation is quantized to $<10,7>$. This ensures that the compressed representation remains both numerically stable and compatible with fixed-point FPGA implementation.
The reconstruction error improves significantly as precision increases from 2 to 8 bits and begins to plateau around 10 bits. This indicates that lower precisions introduce significant information loss, while higher precisions yield negligible performance gains.

Figure~7 additionally shows the distribution of MSE values on the test sample for both the full-precision and quantized models. The close agreement between these distributions demonstrates that 10-bit weight precision provides reconstruction performance nearly identical to that of the full-precision model. This hardware-aware compression drastically reduces the model's complexity with a negligible impact on its ability to reconstruct the original pulse shapes.

The resources that a model uses can be further reduced by adjusting the bit-width and rounding behaviour of the calculations required to emulate the matrix multiplication and sum in the forward pass. 
Two options are presented in this work, both using a fixed precision of $<30,17>$. The ``full'' model (pink line in Fig.~\ref{fig:reco}) implements rounding to the closest representable value and an overflow protection, whereas the ``nano'' model (orange line in Fig.~\ref{fig:reco}) uses truncation instead of rounding and does not have an overflow protection. 
We include both models because the latter provides lower resource utilization, but the former exactly replicates the quantized Python implementation using qKeras, making algorithm performance testing more practical.

\begin{figure}[ht]
  \centering
  \includegraphics[width=0.48\textwidth]{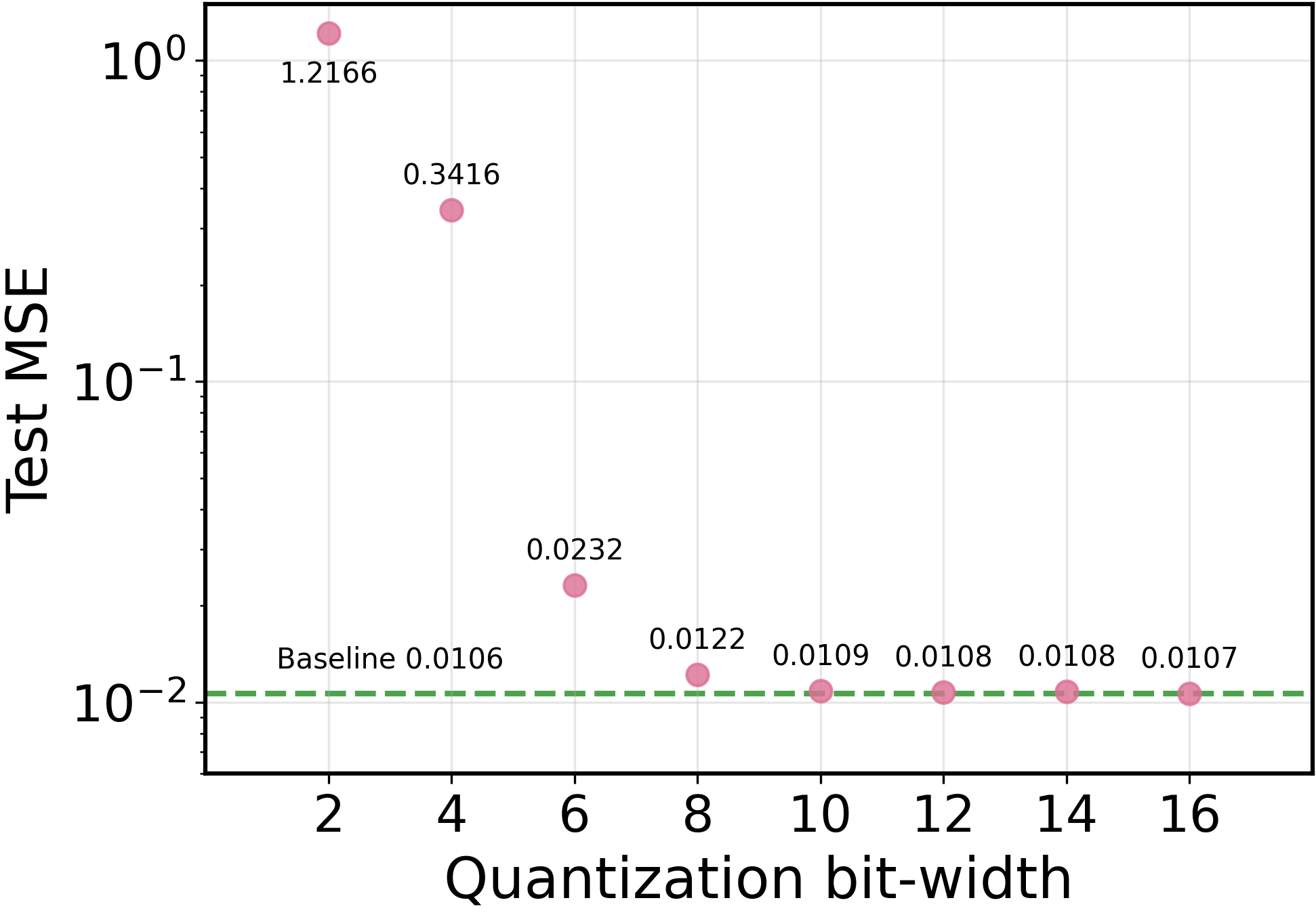}
  \includegraphics[width=0.48\textwidth]{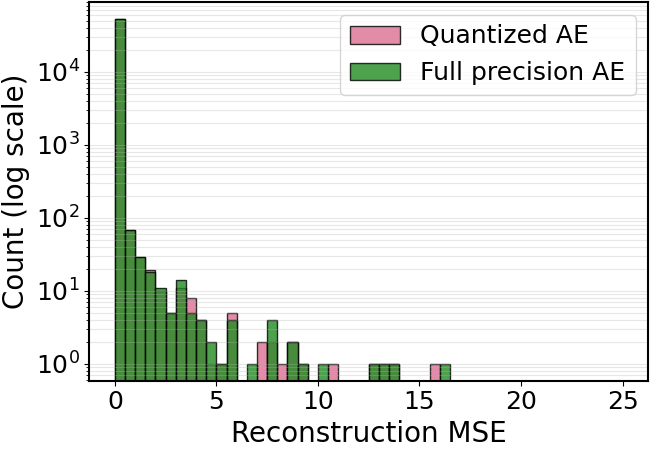}
  \caption{The impact of quantization on the autoencoder's reconstruction performance~(left). The average Mean Squared Error (MSE) is evaluated on the test dataset for different fixed-point bit widths. The green dashed line shows the MSE for non-quantized model. On the right is the distribution of the test MSE for the full precision model vs quantized to 10 bits.}
  \label{fig:mse_vs_quant}
\end{figure}



%% file: sections/hardware_impl.tex
\section{Enabling ML on Radiation-Hard FPGAs: A New \texttt{hls4ml} Backend and Synthesis Results}
\label{sec:hardware_impl}

With an optimized model, the final step is to translate it into firmware for the target FPGA. This required developing new community infrastructure and synthesizing the design to verify its performance against the strict operational constraints of the front-end electronics.

\subsection{A New Backend for \texttt{hls4ml}: Targeting Microchip SmartHLS}
\label{sec:implementation_backend}

The core of our engineering contribution is the development of a new backend for \texttt{hls4ml} that targets the Microchip SmartHLS compiler. This effort effectively bridges the gap between the HEP ML ecosystem and the target hardware platform. The development process involved a systematic translation of the entire compilation flow. To ensure correctness, the development began with a manual C++ implementation to establish a functional baseline for the SmartHLS compiler. While this manual implementation was sufficient for the autoencoder in this study, the primary goal was to create a reusable, automated path for the community. The development of the backend required significant modifications to both the C++ templates and the Python code-generation framework. Key steps included: adapting and optimizing the baseline C++ algorithms to use SmartHLS-native libraries and pragmas; expanding the \texttt{hls4ml} Python data type system to map quantized tensors to the new data types used by SmartHLS (e.g., \texttt{hls::ap\_fixpt}); writing a new code-generation engine tailored to SmartHLS intricacies; and creating the build scripts to integrate the new compiler into the \texttt{hls4ml} workflow. A rigorous validation process was critical to the development, using C-simulations to confirm that the output of the new backend was bit-for-bit identical to that from established \texttt{hls4ml} flows. To ensure a robust hardware verification path, and in coordination with the vendor's technical team, a streaming interface using FIFO buffers was implemented for the top-level module. This design pattern was adopted on their advice to ensure maximum compatibility and to streamline the Software/Hardware co-simulation flow with the current version of the SmartHLS toolchain.

The culmination of this effort was the integration of these changes into a new, fully-fledged \texttt{hls4ml} backend that smoothly integrates into the existing Python API. This new backend, which has been released under  \texttt{hls4ml v1.3.0}, helps democratize access to this class of radiation-hard FPGAs for the scientific community.

\subsection{Synthesis and Performance on PolarFire FPGA}
\label{sec:hardwaredets}

The quantized model was then processed by our new \texttt{hls4ml}-SmartHLS backend to generate HDL code. An important aspect of the hardware implementation relates to how the quantized operations are mapped to the FPGA fabric. The aggressive quantization of the model's weights to a 10-bit fixed-point representation means that the resulting multiplications are implemented efficiently using the general-purpose 4-input LUTs~(Look-Up Tables) rather than the dedicated 18x18 Math Blocks. This behavior, which is consistent with other HLS tools, makes the design highly resource-efficient and underscores the importance of the hardware-aware quantization detailed in Section~\ref{sec:compression}. By tailoring the arithmetic precision, the model avoids relying on specialized DSP resources, which are a limited resource on the FPGA.

This code was subsequently synthesized, placed, and routed for the target Microchip PolarFire MPF100T-FCVG484I and MPF300TS-FCG1152I devices using the Microchip Libero software, yielding the final implementation results. The MPF300TS-FCG1152I provides roughly three times as many LUTs and DFFs as the MPF100T-FCVG484I. Both devices were considered because, although the final FPGA used in PicoCal may be as small as the MPF100T-FCVG484I, the smallest FPGA available in the standard Microchip test kit was the MPF300TS-FCG1152I. We therefore perform the post-synthesis analysis for both FPGAs and verify that the results agree with the behavior observed on the physical MPF300TS-FCG1152I FPGA.

The post-synthesis timing analysis confirms that the design significantly exceeds the requirements, achieving a maximum frequency, $f_{max}$, of 200~MHz (300~MHz) for the FPGA (nano-FPGA) model, well above the target of 160~MHz.
The post-synthesis resource utilization and performance metrics are summarized in Table~\ref{tab:synthesis_results}. The implementation is remarkably efficient, achieving an inference latency between just $25\,$ns for the nano-FPGA model (4 clock cycles at 160~MHz)  and $150\,$ns for the full-FPGA model. With an initiation interval (II) between 3 and 4 clock cycles, the accelerator can process new data every 25~ns, comfortably meeting the 40~MHz data rate requirement of the LHCb front-end. The resource footprint is minimal, consuming $O(1\%)$ of the FPGA's logic (LUTs and Flip-Flops) without using any of its dedicated math blocks. This confirms the solution is not only fast enough for real-time processing but also lightweight enough to be easily integrated alongside other critical logic on the front-end electronics. An example of a resulting floorplan is shown in ~/ref{app:floorplan-example}.

To further validate the new back end, the autoencoder was implemented on the MPF300TS-FCG1152I and tested using a sample of pulses. The resulting output for the less aggressively quantized, i.e. ``full'',  FPGA model, shown by the pink line in Fig.~\ref{fig:reco}, exactly matches the output obtained from the quantized Python implementation. As summarised in  the final row of Table.~\ref{tab:synthesis_results} the results for resource utilisation, as well as timing, were also consistent with the numbers obtained from post-sythesis analysis when the algorithms were implemented on the test kit. The final implementation, including the \texttt{UART} interface used to communicate with the FPGA, used $5578$ ($2995$) LUTs, $6383$ ($2478$) D-Flip-Flops, and $8395$ ($3725$) logic units for the full (nano) FPGA model. 
Finally, the power consumption of the algorithm was assessed using both a centrally provided Microchip prediction tool and the estimates reported by Libero SmartPower. At an ambient temperature of around 20\, degrees~C, both the static power consumption, that is, the power required to keep the FPGA powered on, and the dynamic power, i.e. that required to run the algorithm, are expected to be roughly around $0.1\,$W.


\begin{table}[ht!]
\caption{Post-synthesis resource utilization per channel and performance for the autoencoder model on a Microchip PolarFire MPF100T-FCVG484I and MPF300TS-FCG1152I FPGA.  The utilisation is quoted for the nano-FPGA and full FPGA model using the extrapolation from SmartHLS and the Libero Design suite, which is used to program the FPGA. The target clock frequency for every scenario is 160 MHz (6.25 ns period). The initiation interval (II) and latency are given in units of clock cycles.  The Libero MPF300TS row indicates the performance numbers obtained by running the algorithm on the actual MPF300TS  FPGA, as opposed to via co-simulation. }
\label{tab:synthesis_results}
\begin{tabular}{lcccccc}
\hline
\multicolumn{2}{c}{\multirow{2}{*}{FPGA setup}} & \multicolumn{5}{c}{Value}                                                   \\
\multicolumn{2}{c}{}                     & II & Latency & $f_{\rm max}$ {[}MHz{]} & LUTs                & DFFs                \\ \hline
\multirow{2}{*}{MPF100T}         & FPGA  & 4  & 24      & $180$                   & $7956$ ($7.3\%$)    & $5893$ ($5.4\%$)    \\
                                 & nFPGA & 3  & 4       & $280$                   & $3221$ ($3.0\%$)    & $2059$ ($1.9\%$)    \\
\multirow{2}{*}{MPF300TS}        & FPGA  & 4  & 24      & $212$                   & $7953$ ($2.7\%$)    & $5825$ ($1.9\%$)    \\
                                 & nFPGA & 3  & 4       & $309$                   & $3211$ ($1.1\%$)    & $1861$ ($0.6\%$)    \\
\multirow{2}{*}{Libero MPF300TS} & FPGA  & 4  & 24      & $212$                   & $5578$ ($1.9\%$)    & $6383$ ($2.1\%$)    \\
                                 & nFPGA & 3  & 4       & $309$                   & $2995$ ($1.0\%$)    & $2478$ ($0.8\%$)    \\ \hline
\end{tabular}
\end{table}

%% file: sections/discussion.tex
\section{Discussion}
\label{sec:discussion}

The results presented in the previous sections demonstrate a complete pipeline for ML-based on-detector data compression using FPGAs, from algorithm conception to successful hardware implementation. The synthesis results based on Software/Hardware
co-simulation were found to agree very well with the numbers found when running the algorithm on the actual device. Here, we discuss the implications of these findings, both for the immediate LHCb Upgrade II use case and for the wider community seeking to deploy machine learning in harsh environments.

\subsection{Feasibility for the LHCb Upgrade II Use Case}
A primary goal of this work was to investigate viable solutions for the PicoCal data bandwidth challenge.
It was demonstrated that compressing the signal to two 10-bit numbers preserves the full pulse shape. The rising time and peak amplitude could still be extracted from the compressed pulse, and applying a traditional CFD algorithm to the compressed waveform even yielded improved timestamp resolution compared to the original pulse, due to the smoothing of noise effects. Note that ML-alternatives to the CFD algorithm should be the subject of future work. 

The hardware implementation results confirm that our approach is not only viable but highly effective. The autoencoder model is extremely lightweight, consuming a negligible fraction of the PolarFire FPGA's resources, as shown in Table~\ref{tab:synthesis_results}. This is a critical success factor, as it ensures that the data compression block can be easily integrated onto the front-end electronics without interfering with other necessary functionalities.

Furthermore, the achieved latency and throughput equivalent to the 40~MHz bunch crossing rate meet the demanding real-time requirements of the LHCb trigger and data acquisition system. The successful synthesis proves that a sophisticated ML model can perform a complex task like pulse shape compression well within the allotted time budget. This provides a strong proof-of-principle for the consideration of this implementation by the LHCb collaboration for its future upgrade \cite{LHCb-Upgrade-II-Framework-TDR}.

A crucial aspect of this feasibility study is the scalability of the solution to meet the system's architectural requirements. The resource utilization reported in Table \ref{tab:synthesis_results} corresponds to a single autoencoder instance for one calorimeter channel. As noted, the baseline design for the PicoCal front-end board assumes that each FPGA will process 8 channels. Assuming a parallel implementation where each channel requires a dedicated encoder, the total resource footprint would scale linearly. This would project a total utilization of approximately 25\% of the FPGA's 4-input LUTs ($8 \times 3.10\%$), assuming the nanoFPGA model is running on the smaller MPF100T FPGA.  
This projected usage remains modest and falls comfortably within preliminary resource budgets discussed for this processing task. It leaves a substantial fraction of the FPGA's resources available for other critical functionalities, such as data aggregation and control logic. 


\subsection{The Broader Impact of the \texttt{hls4ml}-SmartHLS Backend}
Perhaps the most significant and lasting contribution of this work is the development of the new \texttt{hls4ml} backend for Microchip's SmartHLS compiler. By creating this piece of open-source infrastructure, we have lowered the barrier to entry for using radiation-hard PolarFire FPGAs for machine learning inference. This tool makes a new class of radiation-hard devices more accessible to the entire high-energy physics community and beyond.

This automated approach provides a significant advantage over a one-off, manual implementation. While the specific autoencoder in this study could have been realized with a manual HLS design, such a solution would lack scalability and reusability. The \texttt{hls4ml} backend, in contrast, empowers the community to rapidly prototype and deploy a wide variety of models without requiring bespoke HLS development for each. This automation was also key to our own design process, enabling the systematic hardware-aware quantization scan presented in Section~\ref{sec:compression}, a crucial optimization step that would have been impractical within a manual design flow.

Experiments at the HL-LHC, future colliders, or in space-based applications can now follow a standard, streamlined workflow to deploy ML models on these robust FPGAs. This accelerates the development cycle and empowers domain experts who are not necessarily FPGA design specialists to implement powerful, on-detector ML solutions. 

It is important to note that the backend, in its current state, supports the specific layers required for this work, namely dense layers and ReLU activations. The immediate plan is to maintain this functionality, providing a robust and stable tool for similar applications. Future development will be driven by community needs, with plans to expand the library of supported layers as new use cases and requirements emerge.

The backend is publicly available and was  integrated into the main library as part of the \texttt{hls4ml} v1.3.0 release in March 2026.

\subsection{Radiation Hardness: A Holistic View}
A key consideration for any electronics deployed in the LHC tunnel is tolerance to radiation-induced errors \cite{lhc-rad-effects}. Our hardware implementation reveals a powerful synergy between our lightweight algorithm and the inherent robustness of the target hardware. The autoencoder model is so resource-efficient (see Table~\ref{tab:synthesis_results}) that its core logic can be physically placed in a radiation-protected region of the FPGA. This leverages the SEU-immune nature of the flash-based configuration memory in the PolarFire FPGA family and benefits from the reduced radiation exposure of this region. Flash-based FPGAs, such as the Microchip PolarFire family used in this work, store their configuration in nonvolatile memory that is inherently immune to configuration SEUs, removing the need for configuration scrubbing or configuration-level TMR required in SRAM-based FPGAs. However, unlike radiation-hardened or space-qualified components, the user logic fabric is not natively protected against SEUs affecting the data path. Registers, state machines, logic elements, and embedded memories remain susceptible to radiation-induced upsets. Therefore, mitigation techniques like TMR or correction techniques may still be required to protect critical functional paths and maintain reliable operation.

In applications where a model is too large for such physical protection, algorithmic mitigation becomes essential. A common approach is a full TMR of the model, which algorithmically strengthens the design against SEUs at the cost of increased resource consumption. In cases where the resource overhead of full TMR is prohibitive, more granular techniques can be applied. For instance, selective persistence approaches, like those enabled by the FKeras framework \cite{fkeras}, can identify the most critical fraction of a model's parameters that require triplication, offering a balance between robustness and resource usage. Alternatively, one could follow fault-aware, quantization-aware training methodologies. These advanced techniques create models that are, by construction, both performant and inherently resilient to faults, making them radiation-hard with minimal redundancy \cite{fault-aware-quantization}. These algorithmic mitigation strategies can further expand the radiation protection offered by the approach described in this paper should a model's resource utilization exceed the capacity of the FPGA's inherently protected logic.

\subsection{Future Directions: Quantifying Physics Performance Gains}

The results presented in this paper provide a strong proof-of-principle for the on-detector compression of calorimeter signals. The ultimate validation, however, will be to quantitatively measure the performance gain achieved by using the compressed information in downstream physics tasks.

A key future study will be to integrate the two-dimensional latent space variables into the clustering algorithms used for reconstructing neutral particles. It is expected that the pulse shape information preserved in the latent space will improve the accuracy with which  energy depositions are associated with their parent particle clusters. This is particularly important for mitigating the effects of increased pile-up, where the ability to disentangle overlapping signals directly impacts the final energy resolution of the measurement.

Such a quantitative analysis was beyond the scope of this design study, primarily because the official clustering algorithms for the LHCb Upgrade~II, which include timing information,  are not yet finalized. The development of these future reconstruction algorithms, which will be designed specifically to exploit precision timing information, is a prerequisite for measuring the full physics impact of the compression strategy demonstrated in this work.

%% file: sections/conclusion.tex
\section{Conclusion}
\label{sec:conclusion}

The increasing data rates and harsh radiation environments of future high-energy physics experiments necessitate the development of novel, on-detector machine learning solutions using FPGAs. This paper presents an end-to-end demonstration of such a system, using the specific data compression challenge of the LHCb Upgrade II PicoCal calorimeter as a testing-ground.

We have delivered a three-fold contribution. First, we developed a lightweight autoencoder that effectively compresses the 32-sample detector pulse shapes into a compact, two-dimensional latent space while preserving the critical information needed for physics reconstruction. Second, we performed a hardware-aware optimization of the model, using quantization to create a highly efficient fixed-point representation with a negligible loss in physics performance. Third, and most significantly, we developed a new backend for the \texttt{hls4ml} library, enabling for the first time the automatic deployment of ML models on radiation-hard Microchip PolarFire FPGAs via the SmartHLS compiler.

Our synthesis results for the target PolarFire FPGA confirm that the implementation is highly efficient. An inference latency as low as 25~ns can be acheived, along with a throughput of 40~MHz, meeting the stringent requirements of the LHCb front-end system. The design also utilizes a minimal fraction of the device's resources. This efficiency allows the model to reside entirely within the FPGA's inherently radiation-hard logic fabric, providing a robust solution for this use case without the need for complex mitigation schemes. 

This work helps  transform a challenging data compression problem into a demonstrated reality. The development of the open-source \texttt{hls4ml}-SmartHLS backend, in particular, provides a lasting contribution to the scientific community, facilitating the adoption of intelligent edge computing in radiation hard environments. Future work will focus on expanding the backend's layer support to broaden its applicability for the community.

%% file: sections/acknowledgements.tex
 The authors would like to thank the LHCb collaboration for providing the context and resources for this design study. We especially thank the Upgrade II team and the Syracuse LHCb group for providing the simulated samples used to train and evaluate the performance of the autoencoder. The work was done with support of NSF Award Number 2411204. We are also grateful for the fruitful discussions with our colleagues within the LHCb data processing team that helped improve this work, including Felipe Souza de Almeida, Nuria Valls Canudas, Christophe Beigbeder, Zulal Kiraz, Christina Agapopoulou, Marina Artuso, Matthew Rudolph, Lauren Mackey, Patrick Robbe and Philipp Roloff. We wish to express our gratitude to the development and support team of Microchip's SmartHLS for their valuable assistance in enhancing the compatibility of the \texttt{hls4ml} library with their framework.